\documentstyle[prb,aps,multicol]{revtex}
\tighten
\begin{document}

\draft
\title{Coulomb blockade of tunneling through a double quantum dot}

\author{K. A. Matveev,$^{(1)}$ L. I. Glazman,$^{(2)}$ and H. U.
  Baranger$^{(3)}$}

\address{$^{(1)}$Massachusetts Institute of Technology, 12-105, Cambridge
  MA 02139\\
$^{(2)}$Theoretical Physics Institute, University of Minnesota,
Minneapolis MN 55455\\
$^{(3)}$AT\&T Bell Laboratories, 600 Mountain Ave. 1D-230, Murray Hill NJ
07974-0636}

\date{Submitted to Phys. Rev. B, 8 December 1995}
\maketitle

\begin{abstract}
  We study the Coulomb blockade of tunneling through a double quantum dot.
  The temperature dependence of the linear conductance is strongly
  affected by the inter-dot tunneling. As the tunneling grows, a crossover
  from temperature-independent peak conductance to a power-law suppression
  of conductance at low temperatures is predicted. This suppression is a
  manifestation of the Anderson orthogonality catastrophe associated with
  the charge re-distribution between the dots, which accompanies the
  tunneling of an electron into a dot. We find analytically the shapes of
  the Coulomb blockade peaks in conductance as a function of gate voltage.

\end{abstract}

\begin{multicols}{2}

\section{Introduction}

Electron tunneling in a mesoscopic structure may be significantly
affected by the charging effects. The charging suppresses tunneling,
if the charge spreading is impeded by weak links, or by a special
geometry of the structure. Such a suppression of tunneling is commonly
referred to as the Coulomb blockade, for a review see
Ref.~\onlinecite{GrabDev}. In recent experiments\cite{Kastner} it has
become possible to observe the Coulomb blockade in semiconductor
heterostructures where the geometry of the system can be easily
modified by adjusting the voltages on special gate electrodes.

A common example of the Coulomb blockade effect is a
measurement of linear conductance between two macroscopic leads weakly
coupled to a quantum dot.\cite{Kastner} When an electron tunnels from a
lead to the dot, the electrostatic energy of the system
\begin{equation}
U=\frac{e^2n^2}{2C}-\kappa enV_g
\label{elenergy}
\end{equation}
changes; here $C$ is the capacitance of the dot, $en$ is its charge, $V_g$
is the gate voltage, and $\kappa= C_g /C$ is a dimensionless geometrical
factor which defines the gate capacitance $C_g$.  At low temperatures,
$T\ll e^2/2C$, the equilibrium discrete charge of the system is determined
by the minimum of $U$. Tunneling of an electron into or out of the dot
leads to a large increase of the energy, and conduction through the dot is
suppressed. However, at certain values of the gate voltage the
electrostatic energy is degenerate,
\begin{equation}
U(n)=U(n+1),
\label{peakcond}
\end{equation}
and the Coulomb blockade is lifted. Therefore, the linear conductance
shows a series of peaks at the gate voltages
$V_g^{*}=(2n+1)e/2C_g$. The heights and shapes of the peaks can
be found\cite{Shekhter} using the master equation technique,
\begin{equation}
G=\frac{G_lG_r}{2(G_l+G_r)}\frac{\kappa e(V_g-V_g^{*})/T}
                                {\sinh[\kappa e(V_g-V_g^{*})/T]}.
\label{sh}
\end{equation}
Here $G_{l,r}$ are the conductances of the weak links connecting the dot
to the leads.

In a number of recent
experimental\cite{Waugh,Molenkamp,Delft,Stuttgart,Munich} and
theoretical\cite{Molenkamp,Ruzin,Sarma,Klimeck,Spectroscopy,Golden} papers
tunneling through two coupled quantum dots was explored. In particular,
by using a double dot structure one can probe the quantum charge fluctuations
more directly than in a single dot.\cite{Waugh,Molenkamp,Spectroscopy}
Here we focus on the geometry of Ref.~\onlinecite{Waugh} shown schematically
in Fig.~\ref{fig:1}, in which the dependence of the peak positions on
the conductance $G_0$ of the constriction between the two dots was
studied.  We discuss the theory of the peak positions in
Sec.~\ref{positions}.  As $G_0$ grows and approaches $2e^2/h$, the peaks
become equidistant, and in this respect the two-dot system behaves as a
single dot of a larger size.  It is clear, however, that unlike a large
single dot, the charge spreading between the two coupled dots is impeded,
and takes a relatively long time, $t\sim C/G_0\sim\hbar(e^2/C)^{-1}$. The
characteristic energy related to this time delay, $\hbar/t$, is of the
order of the charging energy, and one can expect it to affect the
conductance through the double-dot system and cause deviations from
Eq.~(\ref{sh}). Indeed, in Sec.~\ref{shapes} we show that the slow
propagation of charge between the dots results in a suppression of the
conductance peaks. The specific shape and temperature dependence of a
conductance peak provides one with information about quantum fluctuations
of charge between the two dots.

\section{Positions of the peaks in linear conductance}
\label{positions}

To discuss the Coulomb blockade, one has to introduce the electrostatic
energy of the system shown in Fig.~\ref{fig:1}. In the
experiment\cite{Waugh} the potentials of the dots were controlled by a
single gate voltage $V_g$.  Clearly, the equilibrium electrostatic energy
is a function of three variables: the discrete charges of the two dots
$eN_1$ and $eN_2$, and the gate voltage $V_g$. It also depends on the
capacitances of the dots---to the gate, to the external world, and to each
other---which introduce five parameters into the problem. We will use the
following expression for the electrostatic energy
\begin{eqnarray}
U(N_1,N_2) &=& E_C (N_1 + N_2 - 2X)^2
\nonumber \\
    &&+ \tilde{E}_C [N_1 -N_2 +\lambda (N_1 + N_2) - \alpha X]^2,
\label{U}
\end{eqnarray}
where $X$ is a dimensionless variable proportional to $V_g$. The effective
circuit we have in mind, the exact relation between $X$ and $V_g$, and
expressions for the parameters $E_C$, $\tilde E_C$, $\lambda$ and $\alpha$
in terms of the capacitances of the dots can be found in
Appendix~\ref{circuit}.

We intentionally grouped the terms in Eq.~(\ref{U}) in such a way that the
energy depends on the total number of particles in the two dots $N_1+N_2$
and the relative charge $N_1-N_2$. In this paper we assume that the
coupling of the double-dot system to the leads is extremely weak, $G_l,
G_r \ll G_0$, and therefore one can neglect the quantum fluctuations of
$N_1+N_2$.  On the other hand, the inter-dot conductance $G_0$ is not
necessarily small, and at $G_0\sim e^2/h$ the fluctuations of $N_1-N_2$
are significant. The tunneling of the electron between the dots lowers the
ground state energy of the system. Thus to determine the positions of the
peaks in the linear conductance one should generalize Eq.~(\ref{peakcond})
by replacing the electrostatic energy $U$ with the ground state energy of
the double-dot system $E_{\cal N}(X)$ for a fixed total number of
electrons ${\cal N}=N_1+N_2$. That is, the peak positions $X^*$ are given
by
\begin{equation}
E_{\cal N}(X^*) = E_{{\cal N} +1}(X^*).
\label{peakpos}
\end{equation}

Early attempts at the calculation of the peak positions were based on
models allowing only a few discrete states in each
dot.\cite{Sarma,Klimeck} Such an approach should provide an adequate
description of the system in the case of extremely small quantum dots. In
typical experiments,\cite{Waugh,Molenkamp,Delft,Stuttgart,Munich} however,
the number of states in each dot is large, and a model with continuous
spectra of electrons is more appropriate. The calculation of the
ground-state energy for such a model in the limits of weak and strong
coupling between the dots can be found using the techniques developed in
Refs.~\onlinecite{Glazman,Matveev}. For a symmetric system,
$\lambda=\alpha=0$, in the weak tunneling case, $G_0\ll e^2/h$, the peaks
are centered at the following values of the gate
voltage:\cite{Spectroscopy,Golden}
\begin{equation}
X^*_\pm=n+\frac{1}{2}\pm\frac{1}{4}
    \left[1-\frac{\tilde E_C}{E_C}
            \left(1-\frac{2\ln2}{\pi^2}\frac{h G_0}{e^2}\right)\right],
  \label{shifted}
\end{equation}
where $n$ is any integer.
The peak splitting $X_+-X_-$ grows linearly with $G_0$.

In agreement with the experiment,\cite{Waugh} for a symmetric system,
$\alpha=\lambda =0$, the peaks at small conductance $G_0$ are doubly
degenerate (assuming $\tilde E_C=E_C$, which is a good approximation for
the experiment\cite{Waugh}). Even a small asymmetry, $\alpha\ll 1$, lifts
this degeneracy. Indeed, the positions of the peaks as $G_0\to 0$ can be
found from (\ref{peakpos}) with the electrostatic energy (\ref{U}) as the
full energy. As a result we find the two sequences of peaks
\begin{equation}
  X_1^*=\frac{n+1/2}{1+\alpha/2},
  \quad
  X_2^*=\frac{n+1/2}{1-\alpha/2},
\label{sequences}
\end{equation}
where again $n$ is any integer.  An asymmetry of the system caused by a
non-zero $\lambda$ in Eq.~(\ref{U}) also leads to the lifting of the
degeneracy.

One can easily see that the peak positions given by Eq.~(\ref{sequences})
show periodic beats: near certain values of the gate voltage $X$ the
neighboring peaks come very close together---they are separated by a
distance of order $\alpha$---while between those values of $X$ the peaks
are separated by $\delta X\sim1$. The period of these beats is
$\alpha^{-1}$. In the regions where the distance between the neighboring peaks
predicted by Eq.~(\ref{sequences}) is small, an additional splitting due
to the quantum charge fluctuations caused by finite interdot conductance
$G_0$ should be taken into account.  This additional peak splitting can be
found in the same way as the splitting (\ref{shifted}) in the symmetric
case. For the case $\tilde E_C = E_C$, the result is $\delta X =
(\ln2/\pi^2) hG_0/e^2$.

In the opposite case of strong coupling the properties of the system
depend on the particular model of the junction between the dots. For an
electrostatically created constriction between the dots, a one-dimensional
(1D) model of the junction is the most appropriate.\cite{BvH} In this case
the conductance $G_0$ never exceeds $2e^2/h$, and the strong-tunneling
case corresponds to a small reflection coefficient ${\cal R}=1-h
G_0/2e^2\ll1$. We will concentrate on the asymmetric case, $\alpha>0$,
assuming for simplicity $\lambda=0$, and derive the peak positions $X^*$
from Eq.~(\ref{peakpos}). At fixed ${\cal N} =N_1+N_2$ the electrostatic
energy (\ref{U}) can be rewritten as
\begin{equation}
U_{\cal N} (N_1) = E_C({\cal N} -2X)^2 + 4\tilde E_C (N_1 - \gamma)^2,
  \label{U_n}
\end{equation}
where $\gamma=({\cal N} +\alpha X)/2$.  The second term on the right-hand
side is expressed in terms of the number of particles in the left dot.  In
the strong tunneling case $N_1$ is no longer quantized, and at ${\cal
  R}\to1$ its average assumes the value $\langle N_1\rangle=\gamma$, thus
minimizing the electrostatic energy. In this limit one easily finds
$E_{\cal N}(X) = E_C({\cal N}-2X)^2$, and the peaks are equidistant,
$X^*=(2n+1)/4$.

At non-zero ${\cal R}$ the average number of particles in the left dot
$\langle N_1\rangle$ is not precisely equal to $\gamma$, but oscillates
near $\gamma$ with period $\Delta\gamma=1$. The corresponding small
periodic contribution to the ground state energy was found in
Ref.~\onlinecite{Matveev}, where a single quantum dot connected to a large
lead was considered. At temperatures exceeding the level spacings in both
dots the two problems are equivalent, and one can use the
result\cite{Matveev} for the periodic correction to the ground state
energy,
\begin{eqnarray}
E_{\cal N}(X)&=&E_C({\cal N}-2X)^2
\nonumber\\
      & &- \frac{16e^{\bf C}}{\pi^3}{\cal R}\tilde E_C
         \ln\left[\frac{1}{{\cal R}\cos^2\phi(\gamma)}\right]
         \cos^2\phi(\gamma).
  \label{ground_state_energy}
\end{eqnarray}
where ${\bf C}= 0.5772\ldots$ is Euler's constant, and $\phi(\gamma)$ is
defined as $\phi(\gamma)=\phi_0 + \pi\gamma$. In contrast to
Ref.~\onlinecite{Matveev}, we included here a shift $\phi_0$ in the phase
$\phi(\gamma)$ of the Coulomb blockade oscillations. Such a phase shift is
always present, for instance, due to an asymmetry of the scattering
potential in the constriction connecting the dots. In the case of a single
dot connected to a large lead\cite{Matveev} the presence of $\phi_0$ is
irrelevant, as it can always be compensated by an appropriate shift of the
gate voltage. Similarly, in the case of a double-dot system the phase
$\phi_0$ can be incorporated in the definition of $X$ as some shift [see
the definition of $\gamma$ in Eq.~(\ref{U_n})], unless the system is
completely symmetric, $\alpha=0$.

One can now use the expression for the ground state energy
(\ref{ground_state_energy}) to find the corrections to the equidistant
peak positions caused by the weak scattering in the constriction. From
Eq.~(\ref{peakpos}) we find
\begin{equation}
  X^*=\frac{2n+1}{4} +
    (-1)^n \frac{4e^{\bf C}}{\pi^3}\frac{\tilde E_C}{E_C}
    {\cal R}\ln\frac{1}{{\cal R}}\cos(2\phi_0+\pi\alpha X).
  \label{strong}
\end{equation}
This result for the peak positions in the strong tunneling regime, ${\cal
  R}\ll 1$, is a generalization of the results in
Refs.~\onlinecite{Spectroscopy,Golden} to the case of an asymmetric
system. The asymmetry gives rise to the cosine factor in
Eq.~(\ref{strong}). Similarly to the regime of weak tunneling, in the
asymmetric case the distance between the peaks shows beats, with the
period in $X$ being $\alpha^{-1}$.

As we mentioned, the presence of even a weak asymmetry of the system
destroys the periodicity of the peak positions and thus complicates the
comparison of the experimentally observed peak splitting with the
theory.\cite{Spectroscopy,Golden} One should note, however, that in both
weak and strong tunneling cases, in the regions of the gate voltage $X$
where the peak splitting assumes the smallest possible values, the
distance between the neighboring peaks coincides with that predicted by
the theory\cite{Spectroscopy,Golden} for the symmetric case.

In the next section we calculate the heights and shapes of the conductance
peaks, whose positions are given by Eqs.~(\ref{shifted}),
(\ref{sequences}), and (\ref{strong}), and compare the results with the
available experiments.

\section{Heights and shapes of the conductance peaks}
\label{shapes}

\subsection{Weak tunneling between the dots}
\label{weak}

We start our discussion of the heights and shapes of the conductance peaks
with the case of weak tunneling between the dots, which means that the
conductance of the constriction is small, $G_0\ll e^2/h$.  Nevertheless,
we assume that the coupling to the leads is even weaker, $G_l, G_r \ll
G_0$. The results for the conductance depend on the symmetry of the
system. In the symmetric case, one can find the conductance within the
master-equation approach, identical to the one used in a single-dot
case.\cite{Shekhter} The resulting conductance has the form:
\begin{eqnarray}
G&=&\frac{G_l G_r}{G_l + G_r}
    \frac{1}{2+e^{-\beta(X-X_-^*)}
              +e^{-\beta(X_+^*-X)}}
\nonumber\\
 & & \times
     \left[\frac{\beta(X-X_-^*)}
                 {e^{\beta(X-X_-^*)}-1}
           +\frac{\beta(X_+^*-X)}
                 {e^{\beta(X_+^*-X)}-1}
     \right],
\label{nontrivial}
\end{eqnarray}
where $\beta=4E_C/T$, and $X_\pm ^*$ are the positions of the two adjacent
peaks given by Eq.~(\ref{shifted}) with the same $n$. The two peaks are
resolved only at sufficiently low temperatures, $T\ll E_C(X_+^*-X_-^*)$ as
shown explicitly in Fig.~\ref{fig:3}. The derivation of
Eq.~(\ref{nontrivial}) is outlined in Appendix~\ref{master}.

In a symmetric device each state with an odd charge ${\cal N}$ is doubly
degenerate: the ``odd'' electron may be on either the left or right dot.
In addition, at special values of gate voltage $X=X_\pm^*$, states with
charges ${\cal N}$ and ${\cal N}+1$ are degenerate. At these values of
$X$, a charge can be transferred through the double-dot system via a
sequence of real states. As a result, the peak conductance is
temperature-independent.

A small asymmetry changes the situation qualitatively. As we saw in
Sec.~\ref{positions}, the presence of a non-zero $\alpha$ or $\lambda$ in
the electrostatic energy (\ref{U}) lifts the degeneracy; i.e., the states
with an ``odd'' electron on either the left or right dot now have
different energies. If this difference of energies is larger than the
temperature, one can no longer transfer charge through the double-dot
system via real states only. For instance, at the value of the gate
voltage $X_1^*=1/(2+\alpha)$ the energies of the states with charge 0 and
with extra charge $e$ on the left dot are equal, whereas the energy of the
state with charge $e$ on the right dot is $\Delta=4\alpha E_C/(2+\alpha)$.
(Here we assume that all the asymmetry is due to $\alpha>0$, and
$\lambda=0$, $\tilde E_C = E_C$.) At low temperatures the tunneling of an
electron from the left dot to the right one involves an increase of the
energy of the system by $\Delta>T$, and so tunneling through the
double-dot system is suppressed.  Nevertheless, an electron can still
escape from the left dot to the right lead via a virtual state in the
right dot.  Such a mechanism of tunneling in Coulomb blockade devices is
known as {\it cotunneling}.\cite{GrabDev} At temperatures exceeding the
level spacing in the dot, the leading mechanism is inelastic
cotunneling.\cite{Odintsov,Nazarov} In this case an electron tunnels from
the left dot to the right one, and then another electron tunnels from the
right dot to the right lead. After the cotunneling process is completed,
the right dot is returned to the state with no extra charge, but an
electron-hole pair is created. As a result the phase space volume for such
processes is proportional to $T^2$, i.e., the conductance of the system is
now suppressed at low temperatures.

Let us find the height and shape of a conductance peak at low temperatures,
$T\ll \Delta$, for the example described. We first calculate
the rate of cotunneling of an electron from the left dot to the right lead:
\begin{eqnarray}
\frac{1}{\tau}=\frac{2\pi}{\hbar}\sum_{kpqs}&&
               \left|\frac{t_{pk}t_{sq}}{\Delta}\right|^2
               n_k(1-n_p)n_q(1-n_s)
\nonumber\\
 &&\times  \delta(\epsilon_k-\epsilon_p+\epsilon_q-\epsilon_s+\varepsilon),
\label{cotunneling}
\end{eqnarray}
where $t_{pk}t_{sq}/\Delta$ is the second-order matrix element for the
transfer of an electron from state $k$ in the left dot to state $p$ in the
right dot, and then the transfer of another electron from state $q$ in the
right dot to state $s$ in the right lead; $n_{k(p,q,s)}$ and
$\epsilon_{k(p,q,s)}$ are the corresponding Fermi occupation numbers and
energies, respectively. We also defined
\begin{equation}
\varepsilon=U(1,0)-U(0,0)=2(2-\alpha)E_C(X_1^*-X).
\label{varepsilon}
\end{equation}
A straightforward calculation now yields
\begin{equation}
\frac{1}{\tau(\varepsilon)}=
       \frac{\pi\hbar}{3e^4}G_0G_r \left(\frac{T}{\Delta}\right)^2
       \frac{\varepsilon[1+(\varepsilon/2\pi T)^2]}
            {1-e^{-\varepsilon/T}}.
\label{yields}
\end{equation}
Here we used the definition of the conductance $G_0=(2\pi e^2/\hbar)
\sum|t_{pk}|^2\delta(\epsilon_k)\delta(\epsilon_p)$, and a similar relation
for $G_r$.

One can now express the current through the system as
\begin{equation}
I=e\left[\frac{w_1}{\tau(\varepsilon)}
  - \frac{w_0}{\tau(-\varepsilon)}\right],
\label{Current}
\end{equation}
where $w_0$ and $w_1$ are the occupation probabilities of states with the
charge of the left dot 0 and $e$, respectively. Since the escape rate to
the right electrode is strongly suppressed and is much smaller than the
rate of tunneling to the left lead, the left dot is in equilibrium with
the left lead and
\begin{equation}
w_0 = \frac{1}{1+e^{(\varepsilon-eV)/T}},
\quad
w_1 = 1 - w_0.
\label{w}
\end{equation}
Here $V$ is the bias applied to the leads. An expansion of the current
(\ref{Current}) to linear order in $V$ gives the conductance
\begin{equation}
G=\frac{\pi\hbar}{6e^2} G_0 G_r
  \left(\frac{T}{\Delta}\right)^2
  \frac{(\varepsilon/T)[1+(\varepsilon/2\pi T)^2]}
            {\sinh(\varepsilon/T)}.
\label{cotunneling_conductance}
\end{equation}
The dependence $\varepsilon(X)$ is given by Eq.~(\ref{varepsilon}).

As expected the height of the peak is suppressed at low temperatures as
$T^2$. The result (\ref{cotunneling_conductance}) can be applied to any of
the peaks (\ref{sequences}) in the asymmetric system, provided that the
appropriate values of $\Delta$ and $\varepsilon$ are found from the
electrostatic energy (\ref{U}).  The cotunneling peaks calculated for
realistic parameters\cite{Waugh} are presented in Fig.~\ref{fig:4}.

\subsection{The limit of strong tunneling between the dots}
\label{Strong}

In section~\ref{weak} we assumed that the coupling of the two dots is
weak.  This enabled us to apply the standard master equation technique in
the case of a symmetric system, and to account for the lowest-order
cotunneling process only in an asymmetric double-dot device. In this
section we consider the limit of strong tunneling, where the techniques
based on a simple perturbation theory are not applicable. We define the
strong-tunneling limit as the case of perfect transmission through the
channel between the dots, i.e., $G_0=2e^2/h$. To treat this limit, we
apply here a non-perturbative approach based on the bosonized picture of
the 1D transport through the channel.\cite{Flensberg,Matveev,Furusaki}

We shall treat the double dot system as a single conductor of complicated
shape. To find the conductance we will generalize the master-equation
technique of Ref.~\onlinecite{Shekhter} to account for the impedance of
the charge propagation between the dots due to the narrow constriction. We
need to find the renormalization of the rates of tunneling from the leads
into the double-dot conductor. The impedance of the charge redistribution
within this conductor suppresses the tunneling rates, not unlike the
effect\cite{Ingold} of the ``electromagnetic environment'' on transport
through a single tunnel junction.

To find the rate of tunneling through, e.g., the left tunnel junction, we
introduce a Hamiltonian that accounts for electron states in the left lead
and left dot, as well as for the electron states participating in the
re-distribution of the charge between the dots. The separation of the
latter group of states from the two others is possible at time scales
shorter than the time of electron propagation from the tunnel junction to
the other dot. In the case of a single mode constriction, this time is of
the order of the inverse level spacing in the dot. Therefore our theory is
limited to temperatures exceeding the level spacing.

We assume that the constriction connecting the two dots is a single-mode
channel with no reflection. In this case the set of electronic states
responsible for the transport between the dots is one-dimensional and can
be presented in a bosonized form.\cite{Flensberg,Matveev} Thus the
Hamiltonian can be written as $H=H_0+H_C+H_t$,
\end{multicols}
\vspace*{-0.2truein} \noindent \hrulefill \hspace*{3.6truein}
\begin{eqnarray}
H_0&=&\sum_k \epsilon_k a_k^\dagger a_k + \sum_p \epsilon_p a_p^\dagger a_p
      +\sum_{\sigma}\int_{-\infty}^\infty
      \left[\frac{p^2_\sigma}{2mn_0}+\frac{mn_0v_F^2}{2}
      \left(\frac{\partial u_\sigma}{\partial x}\right)^2\right] dx,
\label{H_0}\\
H_C&=& E_C(n_L-2X)^2
  +4\tilde E_C\left\{n_0[u_\uparrow(0)+u_\downarrow(0)]
                     -\frac{1+\lambda}{2}n_L + \frac\alpha2X\right\}^2,
\label{H_C}\\
H_t&=&\sum_{kp}\left(t_{kp}a_k^\dagger a_p F
      + t^*_{kp}a_p^\dagger a_k F^\dagger\right).
\label{tunnel_hamiltonian}
\end{eqnarray}
\begin{multicols}{2}\noindent
First, in $H_0$, since we are considering transport of an electron from
the left lead into the left-hand dot, $a_k$ is the annihilation operator
for electrons in the left lead and $a_p$ is the operator for electrons in
the left dot; $\epsilon_k$ and $\epsilon_p$ are the corresponding
energies. The bosonized 1D electron system is described by the
displacements $u_\sigma(x)$ and momentum densities $p_\sigma(x)$ in two
spin channels, which satisfy the commutation relation $[u_\sigma(x),
p_{\sigma'}(y)] = i\hbar\delta(x-y)\delta_{\sigma\sigma'}$; $m$ and $n_0$
are the mass and density of 1D electrons. Second, in $H_C$, the charging
energy (\ref{U}) is written in terms of the operator $n_L$ of the number
of electrons tunneled through the left barrier and the charge
$en_0[u_\uparrow(0) + u_\downarrow(0)]$ transferred from the left dot to
the right one. Finally, in the tunnel Hamiltonian
(\ref{tunnel_hamiltonian}) the matrix elements $t_{kp}$ describe tunneling
through the barrier. The transfer of each electron into the dot changes
$n_L$ by one; to account for this, we use the operator $F$ defined by the
commutation relation
\begin{equation}
[F, n_L]=F.
\label{commutator}
\end{equation}

The tunneling current through the junction is
\begin{equation}
I_L\equiv e\langle \dot n_L\rangle
                =-\frac{ie}{\hbar}\langle [n_L,H_t]\rangle
                =\frac{2e}{\hbar}{\rm Im}\sum_{kp}t_{kp}^* \langle
                 a_p^\dagger a_k F^\dagger\rangle,
\label{rrate}
\end{equation}
where the average is performed with the density matrix of
the system described by the Hamiltonian
(\ref{H_0})--(\ref{tunnel_hamiltonian}).  Assuming that the transmission
coefficient of the tunnel barrier is small, we will calculate the tunneling
current in lowest (second) order perturbation theory in $t_{kp}$. Thus
we can expand the density matrix up to the first order in $t_{kp}$ and find
\end{multicols}
\vspace*{-0.2truein} \noindent \hrulefill \hspace*{3.6truein}
\begin{equation}
I_L
   = -\frac{2e}{\hbar} {\rm Re}\sum_{kp}|t_{kp}|^2
     \int_{-\infty}^0 dt
     \left[ \langle a_p^\dagger(0)a_p(t)\rangle
      \langle a_k(0)a_k^\dagger(t)\rangle
      \langle F^\dagger(0) F(t)\rangle
     -\langle a_k^\dagger(t)a_k(0)\rangle
      \langle a_p(t)a_p^\dagger(0)\rangle
      \langle F(t) F^\dagger(0)\rangle \right].
\label{rrrate}
\end{equation}
\begin{multicols}{2}\noindent
In thermodynamic equilibrium the two contributions in $I_L$ compensate
each other. To find the effective conductance $G_L$ of the left junction,
which is renormalized due to the slow charge redistribution between the
dots, we now shift the chemical potential in the lead by $eV$ and find
$G_L=dI_L/dV$ in the following form:\cite{Furusaki}
\begin{equation}
G_L = -\frac{i\pi}{\hbar^2} G_l T^2\int_{-\infty}^{\infty}
       \frac{tK(t)dt}{\sinh^2[\pi T(t-i\delta)/\hbar]}.
\label{G_L}
\end{equation}
Here $G_l=(2\pi e^2/\hbar)\sum |t_{kp}|^2
\delta(\epsilon_k)\delta(\epsilon_p)$ is the unrenormalized conductance of
the left barrier, and we have introduced the correlator
\begin{equation}
K(t) = \langle F(t) F^\dagger(0)\rangle.
\label{K(t)}
\end{equation}
In the derivation of Eq.~(\ref{G_L}) we used the equality $\langle F(t)
F^\dagger(0)\rangle= \langle F^\dagger(t) F(0)\rangle$, which follows from
the symmetry of the Hamiltonian (\ref{H_0})--(\ref{tunnel_hamiltonian})
with respect to the transformation $n_L\to-n_L$, $F\to F^\dagger$,
$u_\sigma\to - u_\sigma$, and $X\to -X$.

In the absence of interaction, $E_C=\tilde E_C=0$, the operators $F$ and
$F^\dagger$ commute with the Hamiltonian, the correlator $K(t)=1$, and the
conductance is not renormalized: $G_L=G_l$. We show below that the time
dependence of the correlator $K(t)$ is non-trivial if $\tilde E_C,\, E_C
>0$.  {\it Consequently the effective conductance $G_L$ is renormalized
  and acquires a power-law temperature dependence at $T\ll\tilde E_C$.}

To calculate $K(t)$, we use a unitary transformation $\hat U$ which shifts
the origin of the electron liquid displacement $u_\sigma$ by a distance
which depends on $n_L$,
\begin{eqnarray}
\hat U &=&
\exp\left[i\left(\frac{\alpha}{2} X
                 - \frac{1+\lambda}{2} n_L \right)\Theta
           \right],
\label{unitary}\\
\Theta &=& \frac{1}{2\hbar n_0}\int_{-\infty}^{\infty}
           \bigl[p_\uparrow(y) + p_\downarrow(y)\bigr]dy.
\end{eqnarray}
Upon the transformation (\ref{unitary}) the Hamiltonian is simplified, and
operator $F$ acquires a phase factor:
\begin{eqnarray}
\hat U^\dagger (H_0+H_C)\hat U&=&H_0 +  E_C(n_L-2X)^2
\nonumber\\&&
  +4\tilde E_Cn_0^2[u_\uparrow(0)+u_\downarrow(0)]^2,
\label{transformed}\\
\hat U^\dagger F \hat U &=& F\exp\left(-i\frac{1+\lambda}{2}\Theta\right).
\label{newF}
\end{eqnarray}
The correlation function (\ref{K(t)}) now factorizes, $K=K_FK_\Theta$. The
factor $K_F=\langle F(t) F^\dagger(0)\rangle$ is easily found:
\begin{equation}
K_F = \langle e^{-i E_C(2n_L + 1 -4X)t/\hbar}\rangle
    = \frac{e^{-i4E_C(X^*-X)t/\hbar}}{e^{4E_C(X^* - X)/T} + 1}.
\label{K_F}
\end{equation}
In the derivation of Eq.~(\ref{K_F}) we assumed that the gate voltage $X$
is close to one of the peak positions $X^*=(2n+1)/4$, see
Eq.~(\ref{strong}), and that the temperature is much smaller than $E_C$ so
that only two states are involved. The calculation of $K_\Theta$ is also
straightforward, as the Hamiltonian (\ref{transformed}) is quadratic in
the bosonic variables, and the exponent in (\ref{newF}) is linear in these
variables:
\begin{eqnarray}
K_\Theta(t)
   &=& \left\langle \exp\left[-i\frac{1+\lambda}{2}\Theta(t)\right]
                    \exp\left[i\frac{1+\lambda}{2}\Theta(0)\right]
       \right\rangle
\nonumber\\
   &=& \exp\left\{-\frac{(1+\lambda)^2}{4}
                  \langle[\Theta(0) -\Theta(t)]\Theta(0)\rangle
           \right\}
\nonumber\\
   &=& \left\{\frac{\pi^2 T}{2ie^{\bf C}\tilde E_C}
              \frac{1}{\sinh[\pi T(t-i\delta)/\hbar]}
       \right\}^{(1+\lambda)^2/4}
\label{K_Theta}
\end{eqnarray}

One can now substitute $K=K_FK_\Theta$ into Eq.~(\ref{G_L}) to find the
renormalized conductance,
\begin{equation}
G_L = \frac{G_l}{2}\left(\frac{\pi^2T}{e^{\bf C}
                         \tilde E_C}\right)^{\eta_L^{}}
      F_{\eta_L^{}}\left(\frac{4E_C(X-X^*)}{T}\right),
\label{finalG_L}
\end{equation}
where $\eta_L^{}=(1+\lambda)^2/4$. The peak shape is given by the function
$F_\eta (x)$ defined as
\begin{equation}
F_{\eta}(x)=\frac{1}{\cosh(x/2)}
\frac{\left|\Gamma\left(1+\frac{\eta}{2}+\frac{ix}{2\pi}\right)\right|^2}
{\Gamma(2+\eta)}.
\label{shape}
\end{equation}

The tunneling into the double-dot system is suppressed at low
temperatures, $G_L\propto T^{\eta_L^{}}$. The origin of this suppression
is Anderson's orthogonality catastrophe. The tunneling of an electron into
the left dot results in a significant change of the ground state of the
double-dot system, and the new ground state is orthogonal to the old one,
leading to the suppression of tunneling. Indeed, after the tunneling
process has changed the charge of the left dot by $e$, charge
$q_\uparrow=q_\downarrow=e(1+\lambda)/4$ must be transferred to the right
dot in each spin channel to minimize the electrostatic energy. The
orthogonality of the two ground states results in a power-law suppression
of the tunneling density of states, $G\propto T^{\eta_L^{}}$, where the
exponent can be related to the charges $q_{\sigma}$ as\cite{remark2}
$\eta_L^{}=2\sum_\sigma(q_\sigma/e)^2$, in agreement with
Eq.~(\ref{finalG_L}).

The tunneling through the right barrier can be treated in the same manner,
and the result for the renormalized conductance $G_R$ can be found by
replacement $G_l\to G_r$ and $\eta_L^{}\to \eta_R^{}=(1-\lambda)^2/4$ in
Eq.~(\ref{finalG_L}).  After the renormalized conductances $G_L$ and $G_R$
are found, we can use the master equation approach similar to the one
outlined in Appendix~\ref{master}, and find the total conductance:
\begin{equation}
G=\frac{G_LG_R}{G_L + G_R}.
\label{G}
\end{equation}
At $T\to 0$ the smallest of the two conductances, $G_L$ and $G_R$,
controls $G$, which means that the peak value of $G$ is proportional to
$T^\eta$, with $\eta=(1+|\lambda|)^2/4$. Depending on the geometry of the
system, the parameter $\lambda$ may vary from $-1$ to $1$, and is $0$ in
the symmetric case.  Therefore the exponent of the temperature dependence
$\eta$ varies from $1/4$ in the symmetric case to $1$ in the most
asymmetric case.

\subsection{Intermediate strength of tunneling between the dots}
\label{intermediate}

In the previous two sections we considered the cases of weak and strong
tunneling between the dots. We found that for a symmetric system in the
weak tunneling limit the conductance peak heights are temperature
independent, whereas in the strong tunneling limit the peak heights are
suppressed as $T^{1/4}$. For the asymmetric case we discovered the $T^2$
dependence of the peak conductance in the weak tunneling case, and
$T^\eta$ with geometry-dependent exponent, $\frac14\leq\eta\leq1$, for
strong tunneling. In this section we show that in the intermediate regime
the power-law temperature dependence of the peak conductance persists and
consider the corresponding exponents.

We start with the case of symmetric geometry. The weak tunneling result
(\ref{nontrivial}) was obtained in the framework of the master equation
approach with the inter-dot tunneling rate calculated to first order in
$G_0$. It is known,\cite{Glazman} however, that the higher-order terms of
the perturbation theory give rise to a logarithmic renormalization of the
conductance $G_0$. This renormalization becomes important at temperatures
$T\lesssim T_K\simeq\tilde E_C\exp[-(\pi^3e^2/4\hbar G_0)^{1/2}]$.
Therefore at $G_0\ll e^2/h$ the result (\ref{nontrivial}) is applicable
only in the range of temperatures $T_K\ll T\ll E_C$.

A similar argument can be applied in the vicinity of the strong tunneling
limit. Indeed, it was shown\cite{Matveev} that at $G_0=2(e^2/h)(1-{\cal
  R})$ a weak reflection in the constriction is a relevant perturbation
which grows at low temperatures and becomes strong at $T\lesssim E_C {\cal
  R}$. Thus the $T^{1/4}$-dependence of the peak conductance found in
Sec.~\ref{Strong} holds only in the temperature range $E_C {\cal R}\ll
T\ll E_C$.

To consider the effect of weak backscattering ${\cal R}\ll1$ in the
constriction on the low-temperature asymptotics, $T\ll E_C {\cal R}$, of
the peak conductance, we complement the Hamiltonian
(\ref{H_0})--(\ref{tunnel_hamiltonian}) with a scattering term $H'$. In
bosonic representation $H'$ has the form\cite{Matveev}
\begin{equation}
  H'=-\frac{D}{\pi} \sqrt{\cal R}
     \sum_\sigma\cos[2\pi n_0 u_\sigma(0)-\phi_0],
  \label{H'}
\end{equation}
where the phase shift $\phi_0$ is added to account for the possibility of
an asymmetric location of the scatterer with respect to the center of the
constriction. One can then repeat most of the discussion of
Sec.~\ref{Strong} with the new Hamiltonian. Upon the unitary
transformation (\ref{unitary}) the backscattering term takes the form
\end{multicols}
\vspace*{-0.2truein} \noindent \hrulefill \hspace*{3.6truein}
\begin{eqnarray}
\hat U^\dagger H' \hat U
&=& -\frac{D}{\pi} \sqrt{\cal R}
    \sum_\sigma\cos\left[2\pi n_0 u_\sigma(0)-\phi_X
                         + \frac{\pi}{2}(1+\lambda)n_L
                   \right]
\nonumber\\
&=& -\frac{2D}{\pi} \sqrt{\cal R}
    \cos\left\{\pi n_0[u_\uparrow(0) + u_\downarrow(0)]
           -\phi_X + \frac{\pi}{2}(1+\lambda)n_L\right\}
    \cos\{\pi n_0[u_\uparrow(0) - u_\downarrow(0)]\},
  \label{H'_transformed}
\end{eqnarray}
\begin{multicols}{2}\noindent
where $\phi_X = \phi_0 + \frac{\pi}{2}\alpha X$.

Unlike other terms (\ref{transformed}) of the Hamiltonian, the
backscattering term (\ref{H'_transformed}) shows non-trivial dependence
not only on the sum of the displacements $u_\uparrow + u_\downarrow$, but
also on their difference $u_\uparrow - u_\downarrow$. At low temperatures
$T\ll E_C$, one is only interested in the low-energy behavior of the
system. In this regime the fluctuations of the charge of the dot
$n_0[u_\uparrow(0) + u_\downarrow(0)]$ are frozen due to the charging
energy term in Eq.~(\ref{transformed}) and can be integrated out. The
resulting backscattering term has the form
\begin{eqnarray}
\hat U^\dagger H' \hat U &\simeq&
    -\sqrt{\frac{8e^{\bf C}E_CD{\cal R}}{\pi^3}}
    \cos\left[\phi_X - {\textstyle\frac{\pi}{2}}(1+\lambda)n_L\right]
\nonumber\\
  &&\times  \cos\{\pi n_0[u_\uparrow(0) - u_\downarrow(0)]\},
  \label{H'_integrated}
\end{eqnarray}
cf.\ Ref.~\onlinecite{Matveev}.  Since the operators $F$ and $F^\dagger$
do not commute with the backscattering term (\ref{H'_integrated}), the
latter can affect the $K_F$ component of the correlator $K(t)$ and the
conductance of the left barrier (\ref{G_L}). To find the effect of the
backscattering on $K_F(t)$, we first discuss the influence of the operator
(\ref{H'_integrated}) on the dynamics of the spin field $u_\uparrow -
u_\downarrow$. One can easily show that the operator (\ref{H'_integrated})
is a relevant perturbation,\cite{Furusaki} i.e., the amplitude of the
cosine term grows at low energies. Thus at $T\to 0$ the spin field
fluctuations are frozen at the value $n_0[u_\uparrow(0) -
u_\downarrow(0)]=0$ or 1 for the positive and negative values of
$\cos\left[\phi_X - \frac{\pi}{2}(1+\lambda)n_L\right]$, respectively.

When an electron tunnels into the double-dot system through the left
barrier, the value of $n_L$ changes from 0 to 1. Thus the prefactor in
Eq.~(\ref{H'_integrated}) is proportional to either $\cos\phi_X$ or
$\cos\left[\phi_X - \frac{\pi}{2}(1+\lambda)n_L\right]$. If the two
cosines have the same sign, the increase of $n_L$ described by the
operator $F^\dagger$ does not affect the long-time dynamics of the spin
mode, which remains pinned at the origin with the same value of
$u_\uparrow(0) - u_\downarrow(0)$. In this case the time dependence of
$K_F(t)$ is not affected by the backscattering and the conductance $G_L$
is still given by Eq.~(\ref{finalG_L}), with a different prefactor which
we do not calculate here.  On the other hand, if the signs of $\cos\phi_X$
and $\cos\left[\phi_X - \frac{\pi}{2}(1+\lambda)n_L\right]$ are different,
the change of $n_L$ shifts the boundary condition for the spin mode from
$n_0[u_\uparrow(0) - u_\downarrow(0)]=0$ to 1. An abrupt change of the
boundary condition creates a disturbance in a 1D bosonic field that decays
slowly, giving rise to power-law time dependences of electronic Green's
functions.\cite{Schotte} Thus the correlator $K_F(t)$ acquires an
additional time-dependent factor\cite{Furusaki,remark} $K_s(t)\propto \pi
T/i\sinh[\pi T(t-i\delta)/\hbar]$. According to Eq.~(\ref{G_L}), such a
modification of $K(t)$ not only changes the prefactor in
Eq.~(\ref{finalG_L}), but also replaces the exponent $\eta_L^{}$ by
$\eta_L^{}+1$. Thus depending on the values of $\phi_X$ and $\lambda$, the
backscattering either does not affect the temperature dependence of the
renormalized conductance $G_L\propto T^{(1+\lambda)^2/4}$ or replaces it
with a stronger one, $G_L\propto T^{1+(1+\lambda)^2/4}$.

The latter result can be easily interpreted in terms of the orthogonality
catastrophe. Indeed, as we saw, the tunneling of an electron into the left
dot leads to the transfer of charge $q_\uparrow + q_\downarrow =
e(1+\lambda)/2$ through the constriction. On the other hand,  if
$n_0[u_\uparrow(0) - u_\downarrow(0)]$ changes from $0$ to $1$, the
transferred spin is $(q_\uparrow - q_\downarrow)/2e = \frac12$. Thus we can
easily evaluate the charge transferred in each of the spin channels,
\begin{equation}
q_{\uparrow,\downarrow} = e\left(\pm\frac12 + \frac{1+\lambda}{4}\right).
  \label{charges}
\end{equation}
The suppression of the tunneling density of states is described by the
power law $\nu\propto \epsilon^{\eta_L^{}}$, where the exponent is given
by\cite{remark2} $\eta_L^{} = 2\sum_\sigma(q_\sigma/e)^2$. From
Eq.~(\ref{charges}) we now find
\begin{equation}
\eta_L^{}=1+\frac{(1+\lambda)^2}{4},
  \label{eta}
\end{equation}
which results in the power-law suppression of the conductance, $G_L\propto
T^{1+(1+\lambda)^2/4}$.

To find the temperature dependence of the peaks in the conductance through
the double-dot system, one has to find not only $G_L$, but also $G_R$.  It
is clear that when a tunneling process through the whole double-dot system
is completed, exactly one electron is transferred through the
constriction. Thus we conclude that the total transfer of charge after one
electron has tunneled into the left dot and another one escaped from the
right dot is $\Delta q=e$. Since the charge transferred through the
constriction at the first step was $e(1+\lambda)/2$, the tunneling of an
electron from the right dot must be accompanied by the transfer of charge
$e(1-\lambda)/2$. We saw above that unlike charge, the spin is transferred
in quantized portions $\Delta s = \frac12$. Since the total transferred
spin is $\frac12$, we conclude that exactly one of the two tunneling
events involved the transfer of spin. Therefore in the cases when the
temperature dependence of $G_L$ is given by $T^{(1+\lambda)^2/4}$ and
$T^{1+(1+\lambda)^2/4}$, the conductance $G_R$ behaves as
$T^{1+(1-\lambda)^2/4}$ and $T^{(1-\lambda)^2/4}$, respectively. Finally,
since the total conductance is given by the smaller of $G_L$ and $G_R$, we
find
\begin{equation}
  G \propto
    \begin{cases}
    {T^{1+\frac{(1-\lambda)^2}{4}}, & if $\cos\phi_X
        \cos\!\left[\phi_X - \frac{\pi}{2}(1+\lambda)n_L\right]\!>0$,\cr
     T^{1+\frac{(1+\lambda)^2}{4}}, & if $\cos\phi_X
        \cos\!\left[\phi_X - \frac{\pi}{2}(1+\lambda)n_L\right]\!<0$.}
    \end{cases}
  \label{G_final}
\end{equation}
To determine which option applies to a particular peak, one needs a
detailed knowledge of the microscopic structure of the double-dot system.
It is clear, however, that for nearly symmetric geometries the parameter
$\lambda$, which is determined only by the electrostatics of the system,
should be small: $\lambda\ll1$. {\it In this case we predict the
  temperature dependence $G\propto T^{5/4}$ for all peaks, independent of
  the microscopic structure of the double-dot.}

The temperature dependence of peak heights has been investigated
experimentally by N. van der Vaart {\sl et al}.\cite{Vaart} In the regime
of weak reflection, the power-law fit to the data in the temperature
interval $100{\rm mK}<T<1$K gave $\eta = 0.8$-$1.2$, which is slightly
less than our result $\eta=1.25$ for the symmetric geometry.

The result (\ref{G_final}) shows that the presence of even weak
backscattering in the constriction gives rise to a large correction
$\Delta\eta \sim 1$ to the exponent in the power-law temperature
dependence $G\propto T^\eta$ of the peak conductance. In the derivation of
Eq.~(\ref{G_final}) we assumed that the backscattering is weak, ${\cal
  R}\ll 1$. As the backscattering grows, it further affects the
temperature dependence. Indeed, so far we assumed that the presence of the
backscattering only creates a boundary condition for the spin mode,
$u_\uparrow - u_\downarrow$, and does not affect the charge mode,
$u_\uparrow + u_\downarrow$. However, from the
studies\cite{Glazman,Matveev} of a single dot connected to a large lead it
is known that the backscattering does affect the charge transferred
through the constriction. One can attempt to generalize the result
(\ref{G_final}) to the case of arbitrary ${\cal R}$ by introducing the
value $Q_t$ of the charge transferred through the constriction after an
electron tunnels into the left dot. It is clear from the derivation of
Eq.~(\ref{eta}) that the second term there is actually $Q_t/e$, i.e.,
\begin{equation}
\eta_L^{}=1+\left(\frac{Q_t}{e}\right)^2.
  \label{eta_generalized}
\end{equation}
At ${\cal R}\to 0$ the correction to the electrostatic value
$Q_t=e(1+\lambda)/2$ of the transferred charge is small,\cite{Matveev}
$\Delta Q_t\sim {\cal R}\ln\frac{1}{\cal R}$, which justifies the
approximation (\ref{eta}).

It is interesting to apply Eq.~(\ref{eta_generalized}) to the
weak-tunneling limit ${\cal R}\to 1$, considered in Sec.~\ref{weak}. In
the asymmetric case we demonstrated that the temperature dependence of the
conductance is given by Eq.~(\ref{cotunneling_conductance}). In the
derivation we assumed that because of the high barrier separating the
dots, there is no transfer of electrons between the dots after an electron
tunnels into the left dot. This means that both $Q_t$ and $\Delta s$
vanish, $\eta_L^{} = 0$, and the conductance $G_L$ is not suppressed at
$T\to0$. On the other hand, when an electron escapes to the right lead, it
must go through the constriction, leading to $Q_t=e$ and $\Delta s =
\frac12$. As a result, the relation for $\eta_R$ similar to
Eq.~(\ref{eta_generalized}) will give $\eta_R^{}=2$, which leads to the
quadratic temperature dependence (\ref{cotunneling_conductance}) of the
linear conductance.

In our approach the shape of the peak is obtained from Eq.~(\ref{G_L}) by
substitution $K(t)\propto \{\pi T/i\sinh[\pi T(t-i\delta)/\hbar]\}^\eta$.
As a result the shape of the peak is always uniquely related to its
temperature dependence,
\begin{equation}
G \propto T^\eta F_\eta\left(\frac\varepsilon T\right),
  \label{shape_unique}
\end{equation}
where $F_\eta$ is defined by Eq.~(\ref{shape}), and $\varepsilon$ is
proportional to the deviation of the gate voltage from the peak center.
One can easily check that the peak shape (\ref{cotunneling_conductance})
in the weak-tunneling limit does coincide with $F_2(\varepsilon/T)$.

\section{Conclusions}
\label{conclusions}

In this paper we studied electron tunneling through a system of two
quantum dots connected by a constriction. Tuning the conductance $G_0$ of
this constriction, one may control the quantum charge fluctuations between
the dots and thus affect the Coulomb blockade phenomenon that develops at
a sufficiently low temperature, $T\lesssim E_C$. The positions of the
peaks in linear conductance, $G(V_g)$, depend on the value of $G_0$, and
in the limit $G_0\to 2e^2/h$ the peaks become equidistant
(Section~\ref{positions}). A striking result, however, is that the height
and shape of the peaks also evolve significantly with $G_0$, and remain
non-trivial even in the limit of a reflectionless constriction, $G_0 \to
2e^2/h$.  We have demonstrated that at any $G_0$, except a special case of
small $G_0$ in a symmetric two-dot system (Section~\ref{weak}), the peak
conductance is a power-law function of temperature $T$. The exponent of
the power law depends on the charge re-distribution between the dots that
accompanies the electron transport through the two dots. Both this
exponent and the explicit peak shapes depend on the dots' geometry, as
well as $G_0$ (Sections \ref{Strong} and \ref{intermediate}).  The
suppression of conductance at low temperature and bias can be understood
in quite general terms as an Anderson orthogonality catastrophe caused by
the re-distribution of charge (see
Sections~\ref{Strong},~\ref{intermediate}), and the same exponents should
describe the bias dependence of the differential conductance at $T=0$.

In deriving our results, we assumed that the incoming electron dwells in a
dot for a long time $t_d\gg\hbar/T$ before reaching the constriction that
connects the dots. In a generic situation of a dot lacking a special
symmetry, the electron bounces off the walls many times, before it gets to
the constriction. The dwelling time is determined by the level spacing,
$t_d\sim\hbar/\delta E$. Therefore, the results we presented in
Section~\ref{shapes} are valid in the temperature interval $\delta
E<T<E_C$. For typical parameters, this interval allows one to vary the
temperature by at least one decade.

The effect of quantum charge fluctuations on the ground state energy
has been recently demonstrated experimentally.\cite{Waugh,Molenkamp}
The data of Waugh {\it et al},\cite{Waugh} is in a quantitative agreement
with the present theory.\cite{Spectroscopy} The temperature dependence of
the peak conductance, which is related to the dynamics of the charge
re-distribution, was studied in a very recent experiment by N. van der
Vaart {\it et al}.\cite{Vaart} The temperature dependence exponent found
experimentally in the regime of weak reflection, $\eta = 0.8$-$1.2$, is
somewhat smaller than the theoretical value $\eta=1.25$ we find for this
case.

\acknowledgements

The work at MIT was sponsored by Joint Services Electronics Program
Contract DAAH04-95-1-0038 and at the University of Minnesota by NSF Grant
DMR-9423244.  K.M. and L.G. acknowledge the hospitality of the Aspen
Center for Physics, where this work was completed.
\appendix

\section{Derivation of the electrostatic energy Eq.~(\protect\ref{U})}
\label{circuit}

In this section we find the electrostatic energy of the double dot
structure in terms of the capacitances of the individual dots and the
gate voltage. To describe the electrostatics of the physical structure
shown schematically in Fig.~\ref{fig:1}, we introduce the circuit diagram
in Fig.~\ref{fig:2}. The electrostatics is determined by the gate voltage
$V_g$ and 5 capacitances---$C_1$ and $C_2$ are the capacitances of the
dots to the gate, $C_3$ is the capacitance between the dots, and $C_4$
and $C_5$ are the capacitances of the dots to everything else.

In terms of the charge on each capacitor, the electrostatic energy is
\FL
\begin{equation}
U = \frac{q_1^2}{2C_1} + \frac{q_2^2}{2C_2} + \frac{q_3^2}{2C_3}
  + \frac{q_4^2}{2C_4} + \frac{q_5^2}{2C_5} - q_1 V_g - q_2 V_g.
\label{A:electro1}
\end{equation}
The number of electrons on each dot is given by the sum of the charges
on the appropriate three capacitors:
\begin{eqnarray}
-eN_1 = q_1 + q_3 + q_4,
\\
-eN_2 = q_2 - q_3 + q_5.
\end{eqnarray}
We now must minimize the energy Eq. (\ref{A:electro1}) at fixed values of
$V_g$, $N_1$, and $N_2$ and evaluate the energy at this minimum. The
result has the form (\ref{U}) [up to an irrelevant constant], and we now
give explicit expressions for the parameters in this equation.

First, the energy involved in changing the total charge on the double-dot
system is given simply by the total capacitance of the double-dot to the
external world. Introducing the external capacitance
\begin{equation}
C_{\text{ext}} \equiv C_1 + C_2 + C_4 + C_5,
\end{equation}
we find
\begin{equation}
E_C = \frac{ e^2 } { 2C_{\text{ext}} }.
\end{equation}
The coupling of the total charge to the gate is given by the capacitance
to the gate,
\begin{equation}
    X = \frac{ - ( C_1 + C_2 ) \, V_g } { 2\,e}.
\end{equation}
Turning to asymmetric structures, we find that the fractional asymmetry
of the capacitances determines the parameter $\lambda$,
\begin{equation}
\lambda = ( C_2 + C_5 - C_1 - C_4 )/C_{\text{ext}}.
\end{equation}
In terms of this asymmetry parameter, we find that the charging
energy for transfer from one dot to the other is
\begin{equation}
\tilde{E}_C = \frac{ e^2}
                   {2 [ C_{\text{ext}} (1-\lambda^2) + 4 C_3 ]},
\end{equation}
and that the coupling of this excitation to the gate is given by
\begin{equation}
\alpha = 2 \left( \lambda + \frac{C_1 - C_2}{C_1 + C_2} \right).
\end{equation}
This completely specifies the electrostatic problem.

\section{Master equation technique for the conductance peaks}
\label{master}

In this Appendix we derive the expression (\ref{nontrivial}) for the
conductance peaks in the symmetric case. We restrict ourselves to the case
of temperatures which are much smaller than $E_C$, but can be of the order
of $E_C(X^*_+ - X^*_-)$. For simplicity we assume that the gate voltage
$X$ is close to $\frac12$, so that only the pair of peaks centered at
$X_+^*$ and $X_-^*$ given by Eq.~(\ref{shifted}) with $n=0$ should be
considered. In this regime only the states with charges 0, $e$, and $2e$
should be taken into account. Clearly, one has four states, which can be
denoted as 0, $l$, $r$, and 2, where $l$ and $r$ describe the two states
with charge $e$ on the left and right dot respectively.

We start with introducing the probabilities of occupation of the four
states, which satisfy the obvious condition: $w_0+w_l+w_r+w_2=1$. The rate
of transitions from state 0 to state $l$, which are caused by tunneling of
an electron through the left barrier, is given by
\begin{eqnarray}
\frac{1}{\tau_{0\to l}}
    &=& w_0\frac{2\pi}{\hbar}
       \sum_{kp} |t_{pk}|^2 n_k (1-n_p)
\nonumber\\
    & &\hspace{4em}\times\delta(\epsilon_k + eV+ E_0 - \epsilon_p - E_1)
\nonumber\\
    &=& \frac{G_l}{e^2} w_0 f(E_1 - E_0 - eV).
\label{rate}
\end{eqnarray}
Here $t_{pk}$ is the matrix element of tunneling from the state $k$ in the
left lead to the state $p$ in the left dot, $n_{k(p)}$ and
$\epsilon_{k(p)}$ are the corresponding Fermi occupation numbers and
energies, $E_0(X)$ and $E_1(X)$ are the values of the ground state energy
of the double-dot system with charge $0$ and $1$, the function $f(x)$ is
defined as
\begin{equation}
f(x) = \frac{x}{e^{x/T} -1}.
\end{equation}
In deriving Eq.~(\ref{rate}) we assumed the bias $V$ applied to the left
lead. In a similar way one can find the rates of all other transitions.

In a stationary state the time derivatives of the occupation probabilities
of all the four charge states of the system vanish. This gives us three
independent equations $\dot w_l = \dot w_r = \dot w_2 = 0$, which can be
written in the following form:
\end{multicols}
\begin{eqnarray}
G_l[w_0f(\varepsilon_1-eV) - w_lf(-\varepsilon_1+eV)] + G_0f(0)[w_r - w_l]
 + G_r[w_2f(-\varepsilon_2) - w_lf(\varepsilon_2)] &=& 0,
\label{1}\\
G_r[w_0f(\varepsilon_1) - w_rf(-\varepsilon_1)] + G_0f(0)[w_l - w_r]
 + G_l[w_2f(-\varepsilon_2+eV) - w_rf(\varepsilon_2-eV)] &=& 0,
\label{2}\\
G_l[w_r f(\varepsilon_2-eV) -w_2 f(-\varepsilon_2 + eV)] +
G_r[w_l f(\varepsilon_2) - w_2 f(-\varepsilon_2)] &=& 0.
\label{3}
\end{eqnarray}
Here we have introduced
\begin{eqnarray*}
\varepsilon_1 &\equiv& E_1(X) - E_0(X) = 4E_C(X^*_--X),\\
\varepsilon_2 &\equiv& E_2(X) - E_1(X) = 4E_C(X^*_+-X).
\end{eqnarray*}
The current $I$ can be also expressed in terms of occupation probabilities
$w_i$. In a stationary state, currents through all the junctions are
equal.  Considering the current through the link between the dots, we can
express $I$ in the following form:
\begin{equation}
I=\frac{G_0}{e}f(0)(w_l-w_r).
\label{current}
\end{equation}

The equations (\ref{1})--(\ref{3}) must hold at any bias. Since we are
interested in linear regime $eV\ll T$, we can differentiate equations
(\ref{1})--(\ref{3}) and replace them by the system of equations for the
derivatives $w_0'$, $w_l'$, $w_r'$, and $w_2'$ of the occupation
probabilities over bias:
\begin{eqnarray}
R_l\left(\frac{w_0'}{w_0}- \frac{w_l'}{w_l} + \frac{e}{T}\right)
+ R_0 \left(\frac{w_r'}{w_r}- \frac{w_l'}{w_l}\right)
+ R_{2r}\left(\frac{w_2'}{w_2}- \frac{w_l'}{w_l}\right) &=& 0,
\label{1'}\\
R_r\left(\frac{w_0'}{w_0}- \frac{w_r'}{w_r}\right)
+ R_0 \left(\frac{w_l'}{w_l}- \frac{w_r'}{w_r}\right)
+ R_{2l}\left(\frac{w_2'}{w_2}- \frac{w_r'}{w_r} - \frac{e}{T}\right) &=& 0,
\label{2'}\\
 R_{2l}\left(\frac{w_r'}{w_r}- \frac{w_2'}{w_2} + \frac{e}{T}\right) +
 R_{2r}\left(\frac{w_l'}{w_l}- \frac{w_2'}{w_2}\right) &=& 0,
\label{3'}
\end{eqnarray}
cf.\ Ref.~\onlinecite{Bahlouli}.  Here $w_i$ are the equilibrium
occupation probabilities, and we introduced the equilibrium rates:
\begin{eqnarray}
R_l &=& w_0 G_l f(\varepsilon_1) = w_l G_l f(-\varepsilon_1)
        = G_lf(\varepsilon_1)Z^{-1},
\label{first}\\
R_r &=& w_0 G_r f(\varepsilon_1) = w_r G_r f(-\varepsilon_1)
        = G_rf(\varepsilon_1)Z^{-1},\\
R_0 &=& w_r G_0 f(0) = w_l G_0 f(0) = TG_0e^{-\varepsilon_1/T}Z^{-1},\\
R_{2l} &=& w_r G_l f(\varepsilon_2) = w_2 G_l f(-\varepsilon_2)
        = G_lf(\varepsilon_2)e^{-\varepsilon_1/T}Z^{-1},\\
R_{2r} &=& w_l G_r f(\varepsilon_2) = w_2 G_r f(-\varepsilon_2)
        = G_rf(\varepsilon_2)e^{-\varepsilon_1/T}Z^{-1},
\label{last}
\end{eqnarray}
\begin{multicols}{2}\noindent
with the equilibrium partition function $Z=1+2\exp(-\varepsilon_1/T)+
\exp[-(\varepsilon_1+\varepsilon_2)/T]$. Since the sum of occupation
probabilities always equals one, there are only three independent
variables in the system (\ref{1'})--(\ref{3'}).

Using Eq.~(\ref{current}), we can also express the conductance
\begin{equation}
G=R_0\left(\frac{w_l'}{w_l}-\frac{w_r'}{w_r}\right)
\label{conductance}
\end{equation}
in terms of the solution of the system (\ref{1'})--(\ref{3'}). In the limit
$G_0\gg G_l, G_r$, we find:
\begin{equation}
G=\frac{1}{T}\left(\frac{R_lR_r}{R_l+R_r}+
                  \frac{R_{2l}R_{2r}}{R_{2l}+R_{2r}}\right).
\label{limiting}
\end{equation}
Substitution of the rates (\ref{first})--(\ref{last}) into (\ref{limiting})
yields the formula (\ref{nontrivial}).

\end{multicols}

\begin{figure}
\caption{Schematic view of the double quantum dot system. The dots are
formed by applying negative voltage to the gates (shaded); the solid line
shows the boundary of the 2D electron gas (2DEG).
$V_l$ and $V_r$ create tunnel barriers between the dots and the leads
while $V_0$ controls the transmission coefficient through the constriction
connecting the dots.}
\label{fig:1}
\end{figure}

\begin{figure}
\caption{The evolution of the split peaks with temperature described by
Eq.~(\protect\ref{nontrivial}). The reference point for the gate voltage is
chosen to be $(X_-^*+X_+^*)/2$, and the gate voltage is plotted in units of
$(X_+^*-X_-^*)/2$. The peak splitting is observable at a sufficiently low
temperature,
$\beta \equiv$ $ 4E_C /T \protect\gtrsim 2$.
}
\label{fig:3}
\end{figure}

\begin{figure}
\caption{The conductance as a function of dimensionless gate voltage $X$
in the asymmetric weak-coupling case. Note the correlation between the
modulation of the peak height and the separation of adjacent peaks:
when the peaks are high the splitting is small, while when the peaks are
small they are well separated. The parameters used in
Eq.~(\protect\ref{cotunneling_conductance}) to produce this plot are
$\alpha = 0.155$ and $T/E_C = 0.07$.
}
\label{fig:4}
\end{figure}

\begin{figure}
\caption{Equivalent electrostatic circuit for the double-dot device of
Fig.~\protect\ref{fig:1} in equilibrium.}
\label{fig:2}
\end{figure}


\newpage

\begin{references}

\bibitem{GrabDev}
{\it Single Charge Tunneling,} edited by H. Grabert and M.H. Devoret
(Plenum Press, New York, 1992).

\bibitem{Kastner}
M.A. Kastner, Rev. Mod. Phys. {\bf 64}, 849 (1992);
E.B. Foxman, P.L. McEuen, U. Meirav, N.S. Wingreen, Y. Meir, P.A. Belk,
N.R. Belk, M.A. Kastner, and S.J. Wind,
Phys. Rev. B {\bf 47}, 10020 (1993).

\bibitem{Shekhter}
L.I. Glazman and R.I. Shekhter, J.\ Phys.\ CM {\bf 1}, 5811 (1989).

\bibitem{Waugh}
F.R. Waugh, M.J. Berry, D.J. Mar, R.M. Westervelt, K.C. Campman,
and A.C. Gossard,  Phys. Rev. Lett. {\bf 75}, 705 (1995).

\bibitem{Molenkamp} L.W. Molenkamp, K. Flensberg, and M.  Kemerink,
  Phys. Rev. Lett. {\bf 75}, 4282 (1995).

\bibitem{Delft} N.C. van der Vaart, S.F. Godijn, Y.V. Nazarov, C.J.P.M.
  Harmans, J.E. Mooij, L.W. Molenkamp, and C.T. Foxon, Phys. Rev. Lett.
  {\bf 74}, 4702 (1995).

\bibitem{Stuttgart} R.H. Blick, R.J. Haug, J. Weis, D. Pfannkuche, K. v.
  Klitzing, and K. Eberl, Phys. Rev. B, January 15 (1996).

\bibitem{Munich}
F. Hofmann, T. Heinzel, D.A. Wharam, J.P. Kotthaus, G. B\"ohm, W. Klein,
G. Tr\"ankle, and G. Weimann, Phys. Rev. B {\bf 51}, 13872 (1995).

\bibitem{Ruzin}
I.M. Ruzin, V. Chandrasekhar, E.I. Levin, and L.I. Glazman,
Phys. Rev. B {\bf 45}, 13469 (1992).

\bibitem{Sarma} C.A. Stafford and S. Das Sarma, Phys. Rev. Lett. {\bf
72}, 3590 (1994).

\bibitem{Klimeck} G. Klimeck, G. Chen, and S. Datta, Phys. Rev. B {\bf
50}, 2316 (1994).

\bibitem{Spectroscopy} K.A. Matveev, L.I. Glazman, and H.U. Baranger,
Phys. Rev. B, January 15 (1996).

\bibitem{Golden} J.M. Golden and B.I. Halperin, preprint (5/95).

\bibitem{Glazman} See Sec. 4 of L.I. Glazman and K.A. Matveev, Zh. Eksp.
  Teor.  Fiz.  {\bf 98}, 1834 (1990) [Sov. Phys. JETP {\bf 71}, 1031
  (1990)] and K.A. Matveev, Zh. Eksp. Teor. Fiz. {\bf 99}, 1598 (1991)
  [Sov. Phys. JETP {\bf 72}, 892 (1991)].

\bibitem{Matveev} K.A. Matveev, Phys. Rev. B {\bf 51}, 1743 (1995).

\bibitem{BvH}
See, e.g., C. W. J. Beenakker and H. van Houten in
{\it Solid State Physics}, Vol. 44, edited by H. Ehrenreich and D.
Turnbull (Academic Press, New York, 1991) pp. 109-125.

\bibitem{Odintsov} D.~V. Averin and A.~A. Odintsov, Phys. Lett. A {\bf
    140}, 251 (1989).

\bibitem{Nazarov} D.V. Averin and Yu.V. Nazarov, Phys.\ Rev.\ Lett.\ {\bf
    65}, 2446 (1990).

\bibitem{Flensberg} K. Flensberg, Phys. Rev. B {\bf 48}, 11156 (1993);
Physica B {\bf 203}, 432 (1994).

\bibitem{Furusaki} A. Furusaki and K.A. Matveev, Phys. Rev. B,
  December~15; Phys.  Rev. Lett. {\bf 75}, 709 (1995).

\bibitem{Ingold} G.-L. Ingold and Yu.V. Nazarov, in
  Ref.~\protect\onlinecite{GrabDev}, pp. 21-107.

\bibitem{remark2} This expression for $\eta_L$ follows from the standard
  expression of the exponent for the orthogonality catastrophe $\eta =
  \sum (\delta_m/\pi)^2$ in terms of the phase
  shifts,\cite{Schotte,Nozieres} if one takes into account the Friedel sum
  rule $\delta_m=\pi q_m/e$ and the fact that each spin channel
  corresponds to two semi-infinite 1D modes.

\bibitem{Schotte} K. Schotte and U. Schotte, Phys. Rev. {\bf 182}, 479
(1969).

\bibitem{Nozieres} P. Nozi\`eres and C.T. de Dominicis, Phys. Rev. {\bf
    178}, 1097 (1969).

\bibitem{remark} We have observed a similar effect in Sec.~\ref{Strong},
  where the change of $n_L$ caused a shift of the boundary condition
  for the charge mode and resulted in the contribution (\ref{K_Theta}) to
  $K(t)$.

\bibitem{Vaart} N. van der Vaart, {\it Single Electron Transport and
    Quantum Confinement in Semiconductor Nanostructures}, Thesis, Delft,
  October, 1995.

\bibitem{Bahlouli} H. Bahlouli, K.A. Matveev, D. Ephron, and M.R. Beasley,
  Phys. Rev. B {\bf 49}, 14496 (1994).

\end{references}
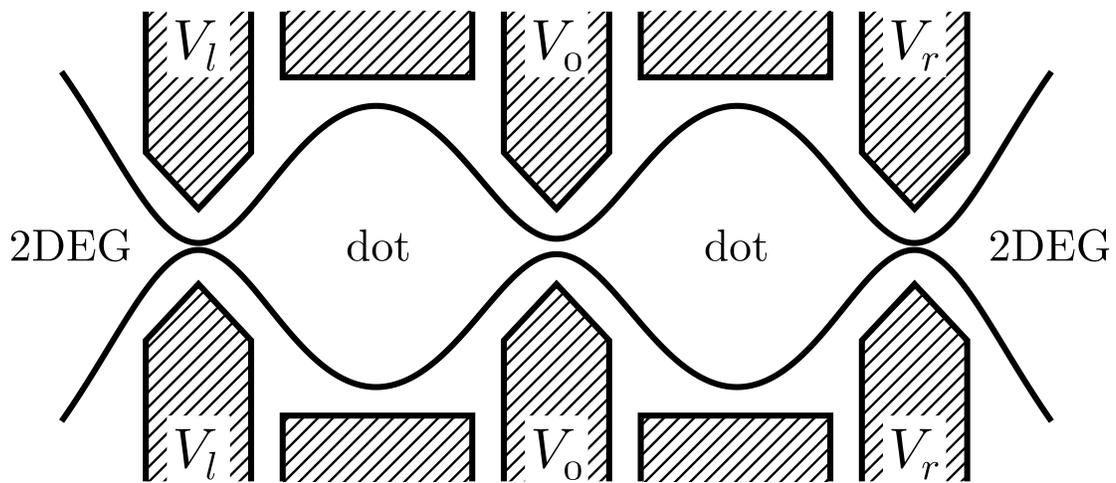
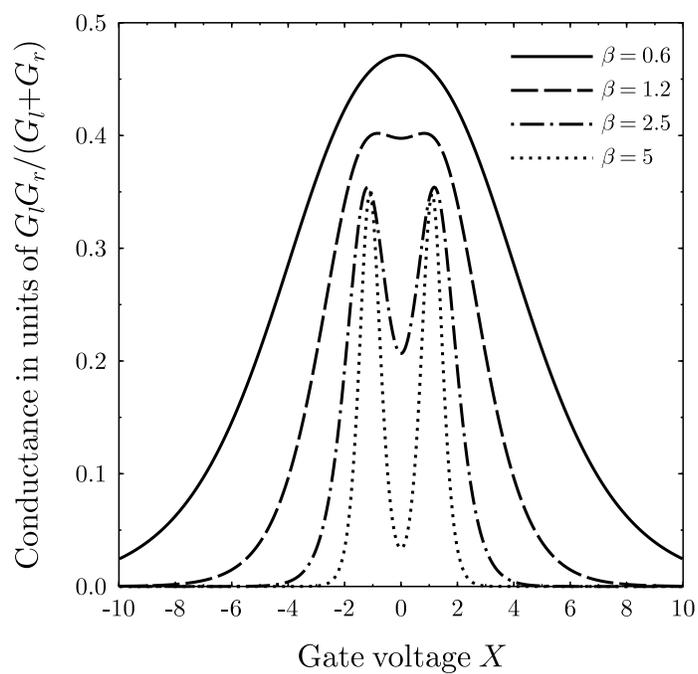
\end{document}